\documentstyle[sprocl]{article}

\input{psfig}

\def\Journal#1#2#3#4{{#1} {\bf #2}, #3 (#4)}

\def\PLB{{\em Phys. Lett.}  B}
\def\PRL{\em Phys. Rev. Lett.}

\def\ZPC{{\em Z. Phys.} C}
\def\EPJC{{\em Euro. Phys. J.} C}

\def\be{\begin{equation}}
\def\ee{\end{equation}}
\def\bea{\begin{eqnarray}}
\def\eea{\end{eqnarray}}
\begin{document}

\title{LEPTOQUARKS AND CONTACT INTERACTIONS \\ FROM A GLOBAL ANALYSIS}

\author{A. F. \.ZARNECKI}

\address{Institute of Experimental Physics, Warsaw University, \\
 Ho\.za 69, 00-681 Warszawa, Poland \\E-mail: zarnecki@fuw.edu.pl }


\maketitle\abstracts{
Data from HERA, LEP and the Tevatron, as well as from low energy  
experiments are used to constrain the scale of possible electron-quark 
contact interactions. 
Some models are found to describe the existing experimental data 
much better than the Standard Model.
%
% The effect is mostly resulting from the new data 
% on the atomic parity violation in cesium,
% but is also supported by recent LEP2 measurements 
% and results concerning CKM matrix unitarity.
%
The possibility of scalar or vector leptoquark contribution
is studied using the Buchm\"uller-R\"uckl-Wyler effective model.
Increase in the global probability observed for scenarios including 
$S_{1}$ or $\tilde{V}_{\circ}$
leptoquark production/exchange  
corresponds to more than a 3$\sigma$ effect.
Assuming that a real leptoquark signal is observed,
calculated is an allowed region in the $\lambda - M$ plane.
 }

\section{Introduction}

In the global contact interaction analysis presented
last year~\cite{gcia,dis99} data from both collider and low energy
experiments were used to constrain the mass scale of 
the possible new electron-quark contact interactions. 
No indication for possible deviations from the Standard 
Model predictions was found at that time.

Presented in this paper are results from the updated analysis, 
which includes new experimental data.
Most important are the new results from the atomic parity violation (APV)
measurements in cesium.\cite{apvnew}
After the theoretical uncertainties have been significantly reduced,
the measured value of the cesium weak charge is 
now 2.5$\sigma$ away from the Standard Model prediction.
Also the new hadronic cross-section measurements at LEP2,
for $\sqrt{s}$=192--202 GeV, are on average about 2.5\% above
the predictions.\cite{lepnew}
This is only about 2.3$\sigma$ effect,
but has an important influence on the analysis.

\section{Contact Interactions}
\label{sec-ci}

Four-fermion contact interactions are an effective theory, which 
allows us to describe, in the most general way, possible low energy 
effects  coming from ``new physics'' at much higher energy 
scales.
Vector $eeqq$ contact interactions can be represented 
as an additional term in the Standard Model Lagrangian:
\begin{eqnarray}
L_{CI} & = & \sum_{i,j=L,R} \eta^{eq}_{ij} (\bar{e}_{i} \gamma^{\mu} e_{i} )
              (\bar{q}_{j} \gamma_{\mu} q_{j})  \nonumber
\end{eqnarray}
where the sum runs over electron and quark helicities.
Couplings  $\eta^{eq}_{ij}$ describing the helicity and flavor 
structure of contact interactions can be related to the effective
mass scale $\Lambda$:
 $\eta = \pm 4 \pi / \Lambda^{2} $.

In the presented analysis different CI scenarios are considered.
The so called first-generation models assume that contact interactions 
couple only electrons to $u$ and $d$ quarks. 
In the three-generation model lepton universality 
($e$=$\mu$) and quark family universality ($u$=$c$=$t$ and 
$d$=$s$=$b$) is assumed.
Assuming $ SU(2)_{L} \times U(1)_{Y} $ gauge invariance,
there are 7 independent couplings ($\eta^{eu}_{RL}$=$\eta^{ed}_{RL}$).
In the most general approach different couplings are allowed to
vary independently, whereas in the so called one-parameter scenarios
only one coupling (or given combination of couplings~\cite{gcia}) 
is allowed to be non-zero.

The analysis combines relevant data from HERA, Tevatron and LEP2, 
results from low-energy $eN$, $\mu N$ and $\nu N$ scattering experiments,
constraints on the CKM matrix unitarity and electron-muon universality,
and the atomic parity violation (APV) measurements.\cite{glqa}

The best description of all data is obtained for three-generation 
model with $e_{L}d_{L}$ type coupling.
Increase in the global probability ${\cal P}(\eta)$ corresponds
to 3.8$\sigma$ deviation from the Standard Model.
The mass scale of new interaction is  $\Lambda^{ed}_{LL} = 13.2 \pm 1.8$ TeV 
(10.3 TeV $< \; \Lambda^{ed}_{LL} \; <$ 21.9 TeV on 95\% CL).

95\% CL exclusion limits on $\eta$ are defined as minimum ($\eta^{-}$) and
maximum ($\eta^{+}$) coupling values resulting in the global probability
equal to 5\% of the Standard Model probability:
${\cal P}(\eta^{\pm}) = 0.05 \; {\cal P}(0)$. 
\begin{table}[t]
\caption{Mass scale limits for different contact interaction models.
         \label{tab:ci}}
\vspace{0.2cm}
\begin{center}
\footnotesize
   \begin{tabular}{|c|rr|rr|rr|rr|}
     \hline
  & \multicolumn{8}{c|}{95\% CL exclusion limit ~~~~ $TeV$} \\ \cline{2-9}
  & \multicolumn{4}{c|}{General approach} & 
    \multicolumn{4}{c|}{One-parameter models} \\ 
\cline{2-9} 
\raisebox{0pt}[8pt]{Coupling}
 & \multicolumn{2}{c|}{1$^{st}$ gen.} & 
               \multicolumn{2}{c|}{3 gen.} &
               \multicolumn{2}{c|}{1$^{st}$ gen.} &
               \multicolumn{2}{c|}{3 gen.} \\
\cline{2-9}
& \raisebox{0pt}[8pt]{$\Lambda^{-}$} & $\Lambda^{+}$ &
  $\Lambda^{-}$ & $\Lambda^{+}$ &
  $\Lambda^{-}$ & $\Lambda^{+}$ &
  $\Lambda^{-}$ & $\Lambda^{+}$ \\ 
\hline
%
 %  Limits from general fits using 1T, 3T models
 %       and 1T and 3T one parameter fits
 %  25/04/2000 12.03.37
 %
 \raisebox{0pt}[8pt]{ $\eta^{ed}_{LL}$} &
 5.3 & 4.4 &
 6.7 & 7.2 &
 28.6 & 8.4 &
 30.5 & 8.9 \\
 \raisebox{0pt}[8pt]{$\eta^{ed}_{LR}$} &
 2.4 & 3.0 &
 3.6 & 3.8 &
 19.2 & 7.6 &
 19.5 & 7.5 \\
 \raisebox{0pt}[8pt]{$\eta$}$^{ed}_{RL}$ &
 3.5 & 3.7 &
 4.3 & 4.2 &
 ~ & ~ &
 ~ & ~ \\
 \raisebox{0pt}[8pt]{$\eta$}$^{ed}_{RR}$ &
 2.8 & 3.5 &
 4.0 & 5.9 &
 8.0 & 18.1 &
 8.8 & 17.4 \\
% \hline
 %
 \raisebox{0pt}[8pt]{$\eta$}$^{eu}_{LL}$ &
 4.9 & 4.6 &
 6.0 & 8.2 &
 12.5 & 18.7 &
 11.8 & 21.4 \\
 \raisebox{0pt}[8pt]{$\eta$}$^{eu}_{LR}$ &
 4.3 & 3.4 &
 5.0 & 4.8 &
 17.4 & 7.5 &
 17.3 & 7.8 \\
 \raisebox{0pt}[8pt]{$\eta$}$^{eu}_{RL}$ &
 3.5 & 3.7 &
 4.3 & 4.2 &
 ~ & ~ &
 ~ & ~ \\
 \raisebox{0pt}[8pt][4pt]{$\eta$}$^{eu}_{RR}$ &
 4.0 & 3.8 &
 4.9 & 4.7 &
 6.9 & 19.7 &
 7.0 & 21.2 \\
\hline
 \raisebox{0pt}[8pt]{$VV$} &
 ~ & ~ &
 ~ & ~ &
 6.5 & 11.1 &
 8.2 & 15.1 \\
 $AA$ &
 ~ & ~ &
 ~ & ~ &
 5.7 & 6.3 &
 10.5 & 11.7 \\
 $VA$ &
 ~ & ~ &
 ~ & ~ &
 4.5 & 4.6 &
 5.7 & 7.8 \\
 %
 %
 % Generated with dis4.kumac
 %

%
\hline
\end{tabular}
\vspace{-3mm}
\end{center}
\end{table}
Mass scale limits $\Lambda^{\pm}$, corresponding the coupling limits
$\eta^{\pm}$, for different contact interaction models
are presented in Table \ref{tab:ci}.

\section{Leptoquarks}

In a recent paper~\cite{glqa} available  data were also 
 used to constrain Yukawa couplings and masses
for scalar and vector leptoquarks
using the Buchm\"uller-R\"uckl-Wyler effective model.\cite{brw}
In the limit of very high leptoquark masses, constraints
on the coupling to the mass ratio were studied using 
the contact-interaction approximation.\cite{lqci}
The best description of the data is obtained for the $S_{1}$ 
and the $\tilde{V}_{\circ}$ leptoquarks~\cite{aachen} with 
$\lambda_{LQ} / M_{LQ} \sim  0.3\; {\rm TeV^{-1}}$.
Increase in the global probability  
corresponds to more than 3$\sigma$ deviation from the Standard Model.

\begin{figure}[t]
\psfig{figure=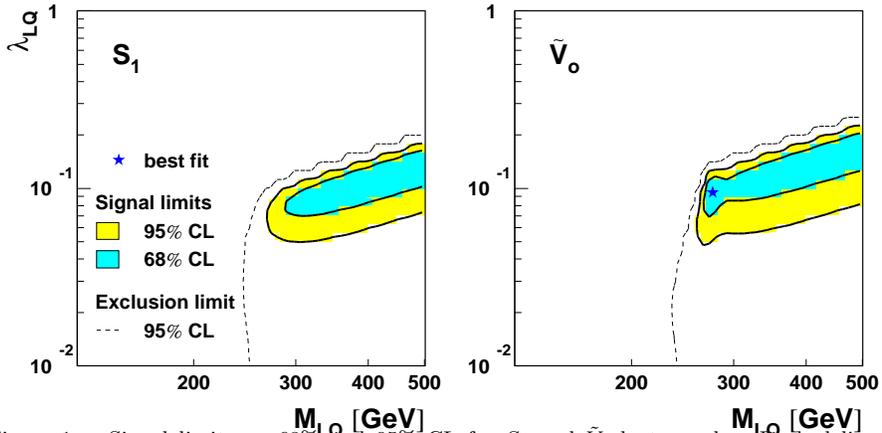,width=\textwidth}
\vspace{-8mm}
\caption{  Signal limits on 68\% and 95\% CL for $S_{1}$ and 
           $\tilde{V}_{\circ}$ leptoquarks. Dashed lines indicate
            the 95\% CL exclusion limits. 
            For $\tilde{V}_{\circ}$ model
            a star indicates the best fit parameters.
            For  $S_{1}$  model the best fit is obtained in the contact
            interaction limit $M_{LQ} \rightarrow \infty$. \label{fig:lq}}
\end{figure}

Constraints on the leptoquark couplings and masses were
studied also for finite leptoquark masses, with mass effects 
correctly taken into account.
Shown in Figure~\ref{fig:lq} are the 95\% exclusion limits
as well as the  68\% and 95\% CL  signal limits for
$S_{1}$ and $\tilde{V}_{\circ}$ leptoquarks.
The best description of the data for the $\tilde{V}_{\circ}$ model
is obtained for $M_{LQ} \; = \; 276 \pm 7$ GeV and
$\lambda_{LQ} \; = \; 0.095 \pm 0.015$.

Table \ref{tab:lq} summarizes the results of the global 
leptoquark analysis.\cite{glqa} For all leptoquark models the 95\% CL
exclusion limits are given both for $\lambda_{LQ}/M_{LQ}$
(upper limit) and for $M_{LQ}$ (lower limit).
For models which describe the existing experimental data 
better than the Standard Model 
the maximum value of the global probability ${\cal P}_{max}$ 
%  (normalised to the Standard Model probability)
and  the corresponding coupling to the
mass ratio $\left( \lambda_{LQ}/M_{LQ} \right)_{max}$ 
are included.
%
% For models with  ${\cal P}_{max}>20$ signal limits on 95\% CL,
% corresponding to ${\cal P}(\lambda_{LQ},M_{LQ}) = 0.05 \;{\cal P}_{max}$, 
% are given for  $\lambda_{LQ}/M_{LQ}$ and $M_{LQ}$. 
%
95\% CL signal limits  for  $\lambda_{LQ}/M_{LQ}$ and $M_{LQ}$,
defined by the condition 
${\cal P}(\lambda_{LQ},M_{LQ}) > 0.05 \;{\cal P}_{max}$, 
are given for models with  ${\cal P}_{max}>20$.
\begin{table}[t]
\caption{ Results of the global leptoquark analysis:
     the 95\% CL  exclusion limits on the leptoquark coupling to the
     mass ratio $\lambda_{LQ}/M_{LQ}$ (upper limit) 
     and the leptoquark mass $M_{LQ}$ (lower limit),
    the coupling to the mass ratio
    $( \lambda_{LQ}/M_{LQ} )_{max}$
   resulting in the best description of the experimental data
   and the corresponding model probability ${\cal P}_{max}$,
   and  the 95\% CL  signal limits on $\lambda_{LQ}/M_{LQ}$ and $M_{LQ}$,
   for models with ${\cal P}_{max}>20$. 
   Global probability function is defined in such a way that 
   the Standard Model probability ${\cal P}_{SM} \equiv 1$.
   \label{tab:lq}}
\vspace{0.2cm}
\begin{center}
\footnotesize
   \begin{tabular}{|l|cc|cc|cc|}
      \hline
      &  \multicolumn{2}{c|}{95\% CL }
      &  \multicolumn{2}{c|}{best description}
      & \multicolumn{2}{c|}{95\% CL  } \\ \cline{4-5}
      &  \multicolumn{2}{c|}{excl. limits}
      &    & 
      & \multicolumn{2}{c|}{signal limits } \\ \cline{2-3} \cline{6-7}
 Model &  \raisebox{0pt}[11pt][7pt]{$\frac{\lambda_{LQ}}{M_{LQ}}$} 
       & $M_{LQ}$
      &  \raisebox{4pt}[-8pt][-8pt]{$\left(\frac{\lambda_{LQ}}{M_{LQ}}\right)_{max}$}
      & ${\cal P}_{max}$     
      &  \raisebox{0pt}[-8pt][-8pt]{$\frac{\lambda_{LQ}}{M_{LQ}}$} 
      & $M_{LQ}$ \\
      & $TeV^{-1}$   & $GeV$ & $TeV^{-1}$  & &$TeV^{-1}$  &  $GeV$ \\ 
\hline
%
%
%  Leptoquark limits in CI approximation
%           20/04/2000  15.48.00
%
\raisebox{0pt}[8pt]{ $S_{\circ}^L$}  &  0.27 & 213 &     &   &   &   \\  
$S_{\circ}^R$  &  0.25 & 242 &     &   &   &   \\
$\tilde{S}_{\circ}$  &  0.28 & 242 &   &   &   &  \\
$S_{1/2}^L$    &  0.29 & 229 &      &   &   &  \\
$S_{1/2}^R$    & 0.49  & 245 & 0.32 $\pm$ 0.06 &  35.8 & 0.09--0.44 & 258 \\  
$\tilde{S}_{1/2}$  & 0.26 & 233 &    &   &   &  \\  
$S_1$  & 0.41 & 245 & 0.28 $\pm$ 0.04 &  367. & 0.15--0.36 &  267 \\  
\hline
\raisebox{0pt}[8pt]{ $V_{\circ}^L$}  &  0.12 & 230 &     &   &   &  \\ 
$V_{\circ}^R$  & 0.44 & 231 & 0.28 $\pm$ 0.07 &  11.7 &  & \\ 
$\tilde{V}_{\circ}$ & 0.52 & 235 & 0.34 $\pm$ 0.06 & 122. & 0.16--0.46 & 259 \\  
$V_{1/2}^L$  & 0.47 & 235 & 0.30 $\pm$ 0.06 &  31.7 & 0.08--0.42 &  254 \\  
$V_{1/2}^R$  &  0.13 & 262 &         &    &  &  \\
$\tilde{V}_{1/2}$  & 0.47 & 244 & 0.30 $\pm$ 0.07 &  14.8 & &  \\
$V_1$  &  0.14 & 254 &    &   &   &  \\
\hline
\end{tabular}
\end{center}
\end{table}

\section*{Acknowledgments}
This work has been partially supported by the Polish State Committee 
for Scientific Research (grant No. 2 P03B 035 17).


\section*{References}
\begin{thebibliography}{99}

\bibitem{gcia}   % My global analysis
A.F. \.Zarnecki \Journal{\EPJC}{11}{539}{1999}. 

\bibitem{dis99}   % My DIS99 contribution
A.F. \.Zarnecki \Journal{\em Nucl. Phys. Proc. Suppl.}{79}{158}{1999}.

\bibitem{apvnew}    % New APV results
S.C. Bennett and C.E. Wieman, \Journal{\PRL}{82}{2484}{1999}.

\bibitem{lepnew}    % LEP new results
LEP Electroweak Working Group, C.Geweniger {\it et al}, LEP2FF/00-01.

\bibitem{glqa}   % My global LQ analysis
A.F. \.Zarnecki hep-ph/0003271.

\bibitem{brw}   % BRW model
W.Buchm\"uller, R.R\"uckl and D.Wyler, \Journal{\PLB}{191}{442}{1987}; \\
Erratum: \Journal{\PLB}{448}{320}{1999}.

\bibitem{lqci}  % LQ in CI approximation
J.Kalinowski {\it et al},    \Journal{\ZPC}{74}{595}{1997}.

\bibitem{aachen} % Aachen notation
A.Djouadi, T.K{\"o}hler, M.Spira, J.Tutas,  \Journal{\ZPC}{46}{679}{1990}.

\end{thebibliography}
\end{document}